\documentclass[12pt]{article}
\newcommand{\bfi}{\bfseries\itshape}
\newcommand{\singlespace}{\baselineskip 4.2333mm\parskip0.3mm}
\def\setmarsing-cm{
\voffset-2.0cm
\hoffset-1.4cm
\textwidth17.0cm
\textheight22.5cm}

\def\setmarsing{in}{
\voffset-.8in
\hoffset-.6in
\textwidth6.5in
\textheight9in}

\def\setmarsing-wide{
\textwidth 7in
\textheight 9.0in
\oddsidemargin -0.25in
\topmargin -0.75in}

\def\setmarsing-ok{
\textwidth 6.5in
\textheight 8.5in
\oddsidemargin 0in
\topmargin -0.5in}

\def\frac#1#2{{{#1}\over {#2}}}

\setmarsing-ok

\begin{document}

\title{{\bf Navier-Stokes-alpha model: LES
equations  with nonlinear dispersion}}
\author{
{\small J. A. Domaradzki}
\\{\small Department of Aerospace }
\\{\small and Mechanical Engineering}
\\{\small University of Southern California}
\\{\small Los Angeles, CA 90089-1191}
\\{\footnotesize email: jad@usc.edu}
\and
{\small Darryl D. Holm}
\\{\small T-Division and CNLS, MS-B284}
\\{\small Los Alamos National Laboratory}
\\{\small Los Alamos, NM 87545, USA}
\\{\footnotesize email: dholm@lanl.gov}
}
\date{\hfil
Special LES volume of ERCOFTAC bulletin, To appear 2001}

\maketitle


\begin{abstract}\noindent
We present a framework for discussing LES equations with
nonlinear dispersion. In this framework, we discuss the properties of
the nonlinearly dispersive Navier-Stokes-alpha (NS$-\alpha$) model
of incompressible fluid turbulence --- also called the viscous
Camassa-Holm equations in the literature --- in comparison with the
corresponding properties of large eddy simulation (LES) equations
obtained via the approximate-inverse approach.
In this comparison, we identify the
spatially filtered NS$-\alpha$ equations with a class of generalized LES
similarity models. Applying a certain approximate inverse to this class
of LES models restores the Kelvin circulation theorem for the defiltered
velocity and shows that the NS$-\alpha$ model describes the dynamics of
the defiltered velocity for this class of generalized LES similarity
models. We also show that the subgrid scale forces in the NS$-\alpha$
model transform covariantly under Galilean transformations and under
a change to a uniformly rotating reference frame. Finally, we
discuss in the spectral formulation how the NS$-\alpha$ model retains
the local interactions among the large scales\,; retains the nonlocal
sweeping effects of large scales on small scales\,; yet attenuates the
local interactions of the small scales amongst themselves.

\end{abstract}


\section{Introduction}
The effects of subgrid-scale (SGS) fluid motions occurring below the
available resolution of numerical simulations must be modeled.
One way of modeling these effects is simply to discard the energy
that reaches such subgrid scales. This is clearly unacceptable,
though, and many creative alternatives have been offered. A
prominent example is the large eddy simulation (LES) approach,
see, e.g., \cite{MGermano[1992]} - \cite{SandipGhosal[1999]}.
The LES approach is based on applying a spatial filter to the
Navier-Stokes equations. The reduction of flow complexity and
information content achieved in the LES approach depends on the
characteristics of the filter that one uses, its type and width.
In particular, the LES approach introduces a length scale into
the description of fluid dynamics, namely, the width of the
filter used. The LES approach is conceptually different from the
Reynolds Averaged Navier-Stokes (or, RANS) approach, which is based on
statistical arguments and exact ensemble averages, rather than spatial
and temporal filtering. After filtering, however, just as in the RANS
approach, one faces the classic turbulence closure problem: How to model
the effects of the filtered-out subgrid scales in terms of the remaining
resolved fields? In practice, this problem is compounded by the
requirement that the equations be solved numerically, thereby
introducing further approximations.

A turbulence modeling scheme --- called here the Navier-Stokes-alpha
model, or NS$-\alpha$ model (also called the viscous Camassa-Holm
equations in \cite{Chen-etal-PRL[1998]} - \cite{FHT-JDDE[2000a]}) ---
was recently introduced. This model imposes an energy ``penalty"
inhibiting the creation of smaller and smaller excitations below a
certain length scale (denoted alpha), but still allows for nonlinear
sweeping of the smaller scales by the larger ones. This energy penalty
implies a {\bfi nonlinearly dispersive modification} of the
Navier-Stokes equations. The alpha-modification appears in the
nonlinear convection term, it depends on length scale and we emphasize
that it is dispersive, not dissipative. This modification causes the
translational kinetic energy wavenumber spectrum of the NS$-\alpha$
model to roll off rapidly below the length scale alpha as $k^{-3}$ in
three dimensions, instead of continuing to follow the slower Kolmogorov
scaling law, $k^{-5/3}$, thereby shortening its inertial range and
making it more computable \cite{FHT-PhysD[2000b]}.

This roll-off in the energy spectrum means that the inertial range is
curtailed and the length scale $\ell_\alpha$ at which viscous dissipation
takes over in the NS$-\alpha$ model is {\it larger} than for the
Navier-Stokes equations with the same viscosity. Hence, for a
given driving force and viscosity, the number of active degrees of
freedom $N_{dof}^\alpha$ for the NS$-\alpha$ model is {\it smaller} than for
Navier-Stokes. A rigorous analytical estimate  was derived in
\cite{FHT-JDDE[2000a]} for the fractal dimension $D_{frac}$ of the {\it
global attractor} for the NS$-\alpha$ model. Namely,
\begin{equation}\label{frac-dim-alpha}
D_{frac} \leq (N_{dof}^\alpha)^{3/2}
\quad\hbox{and}\quad
N_{dof}^\alpha
\equiv
 (L/\ell_\alpha)^3
\simeq
 \frac{L}{\alpha}Re^{3/2}
\,,
\end{equation}
where $L$ is the integral scale (or domain size), $\ell_\alpha$
is the end of the NS$-\alpha$ inertial range and
$Re=L^{4/3}\varepsilon_\alpha^{1/3}/\nu$ is the Reynolds number
(with NS$-\alpha$ energy dissipation rate $\varepsilon_\alpha$ and
viscosity $\nu$).  The Kolmogorov dissipation length scale is
$\ell_{Ko}= (\nu^3/\varepsilon_\alpha)^{1/4}$. The three length scales
$\alpha>\ell_\alpha>\ell_{Ko}$ are related by
\begin{equation}
\ell_\alpha^{\,3}
\simeq
\alpha\,\ell_{Ko}^{\,2}
\quad\hbox{or}\quad
\frac{\alpha}{\ell_\alpha}
\simeq
\frac{\ell_\alpha^{\,2}}{\ell_{Ko}^{\,2}}
\,.
\end{equation}
Thus, these length scales $\alpha>\ell_\alpha>\ell_{Ko}$ stand in
relation to each other in the same way as the length scales
$(K^3/\varepsilon^2)^{1/2}
>(K\nu/\varepsilon)^{1/2}>(\nu^3/\varepsilon)^{1/4}$ in a K-epsilon
model of Navier-Stokes turbulence.

The number of degrees of freedom for a corresponding Navier-Stokes flow
with the same viscosity and energy dissipation rate is
\begin{equation}
N_{dof}^{NS}
\equiv
 (L/\ell_{Ko})^3
\simeq
Re^{9/4}
\,,
\end{equation}
The implication of these estimates of degrees of
freedom for direct numerical simulations that access a significant
number of them using the NS$-\alpha$ model is an increase
in computational speed relative to Navier-Stokes of
\begin{equation}\label{factor}
\bigg(\frac{N_{dof}^{NS}}{N_{dof}^\alpha}\bigg)^{4/3}
\simeq
 \bigg(\frac{\alpha}{L}\bigg)^{4/3}
\!\! Re\,.
\end{equation}
Thus, if $\alpha$ tends to a constant value, say $L/100$, when the
Reynolds number increases -- as found in
\cite{Chen-etal-PRL[1998]} - \cite{Chen-etal-PhysD[1998]} by
comparing steady NS$-\alpha$ solutions with experimental
data for turbulent flows in pipes and channels -- then one could
expect to obtain a substantial increase in computability by using
the NS$-\alpha$ model at high Reynolds numbers. An early
indication of the reliability of using these estimates to gain a
relative increase in computational speed in direct numerical
simulations of homogeneous turbulence in a periodic domain is
given in \cite{Chen-etal[1999]}.

In this paper, we introduce the NS$-\alpha$ model as one of a class of
equations obtained by filtering the Kelvin circulation theorem for
Navier-Stokes. These equations possess the vortex stretching and
transport properties emphasized in \cite{FHT-PhysD[2000b]}. We then
compare a spatially filtered version of the NS$-\alpha$ equations with a
generalization of the LES similarity models.
We also calculate the kinematic tranformation properties of the
NS$-\alpha$ model under changes of reference frames and discuss its
nonlinear spectral-mode dynamics.


\section{Kelvin-filtered turbulence models} Following
\cite{FHT-PhysD[2000b]}, the Navier-Stokes-alpha (NS$-\alpha$) model may
be introduced by starting from Kelvin's circulation theorem for the
original Navier-Stokes (NS) equations,
\begin{equation}\label{NS-eqns}
\frac{\partial\mathbf{v}}{\partial t}
+ \mathbf{v}\cdot\nabla\mathbf{v} + \nabla p
= \nu\nabla^2\mathbf{v}
+ \mathbf{f}\,,
\quad\hbox{with}\quad
\nabla\cdot\mathbf{v}
= 0
\,,
\end{equation}
for an appropriate forcing $\mathbf{f}$ and constant kinematic viscosity
$\nu$. These equations satisfy {\bfi Kelvin's circulation
theorem},
\begin{equation}\label{KelThm-NS}
\frac{d}{dt}
\oint_{\gamma(\mathbf{v})}
\mathbf{v}\cdot d\mathbf{x}
=
\oint_{\gamma(\mathbf{v})}
(\nu\nabla^2\mathbf{v}
+ \mathbf{f})\cdot d\mathbf{x}
\,,
\end{equation}
for a fluid loop $\gamma(\mathbf{v})$ that moves with velocity
$\mathbf{v}(\mathbf{x},t)$, the Eulerian fluid velocity.


\paragraph{Kelvin-filtering the Navier-Stokes equations.}
The equations for the NS$-\alpha$ model may be introduced by modifying
the NS Kelvin circulation theorem (\ref{KelThm-NS}) to integrate
around a loop $\gamma(\hat{\mathbf{v}})$ that moves with a {\bfi
spatially filtered Eulerian fluid velocity} given by
$\hat{\mathbf{v}}=g*\mathbf{v}$, where $*$ denotes the convolution,
\begin{equation}\label{g-star}
\hat{\mathbf{v}}=g*\mathbf{v}
= \int g(\mathbf{x}-\mathbf{x}^{\,\prime\,})
\,\mathbf{v}(\mathbf{x}^{\,\prime\,})\
d^{\,3}x^{\,\prime}
\equiv L_g\mathbf{v}(\mathbf{x})
\,.
\end{equation}
The ``inverse'' operation is denoted
\begin{equation}\label{g-inv}
\mathbf{v}=\mathcal{O}\hat{\mathbf{v}}
\equiv L_g^{-1}\hat{\mathbf{v}}
\,,
\end{equation}
thereby defining an operator $\mathcal{O}$ whose Green's
function is the filter $g$ and which we shall assume is
positive, symmetric, isotropic, translation-invariant and
time-independent. Under these assumptions the quantity (kinetic energy)
\begin{equation}\label{KE-norm}
E =
\frac{1}{2}\int
\mathbf{v}\cdot\hat{\mathbf{v}}\ d^3x
=
\frac{1}{2}\int \mathbf{v}\cdot g*\mathbf{v}\
d^3x
= \frac{1}{2}\int\hat{\mathbf{v}}\cdot\mathcal{O}\hat{\mathbf{v}}\
d^3x
\,,
\end{equation}
defines a norm.

We modify the Navier-Stokes equations (\ref{NS-eqns}) by replacing in
their Kelvin's circulation theorem (\ref{KelThm-NS}) the loop
$\gamma(\mathbf{v})$ with another loop $\gamma(\hat{\mathbf{v}})$ moving
with the spatially filtered velocity, $\hat{\mathbf{v}}$. Then we have,
\begin{equation}\label{KelThm-NS-alpha}
\frac{d}{dt}
\oint_{\gamma(\hat{\mathbf{v}})}
\mathbf{v}\cdot d\mathbf{x}
=
\oint_{\gamma(\hat{\mathbf{v}})}
(\nu\nabla^2\mathbf{v}
+ \mathbf{f})\cdot d\mathbf{x}
\,.
\end{equation}
After taking the time derivative inside the Kelvin loop
integral moving with filtered velocity $\hat{\mathbf{v}}$ and
reconstructing the gradient of pressure, we find the
{\bfi Kelvin-filtered Navier-Stokes equation},
\begin{equation}\label{NS-alpha-eqns}
\frac{\partial\mathbf{v}}{\partial t}
+ \hat{\mathbf{v}}\cdot\nabla\mathbf{v}
+ \nabla\hat{\mathbf{v}}^{\rm T}\cdot\mathbf{v}
+ \nabla p
= \nu\nabla^2\mathbf{v}
+ \mathbf{f}\,,
\end{equation}
with
\begin{equation}\label{aux-cond}
\nabla\cdot\hat{\mathbf{v}}
= 0
\,,\quad\hbox{and}\quad
\mathbf{v}=\mathcal{O}\hat{\mathbf{v}}
\,.
\end{equation}
The transport velocity $\hat{\mathbf{v}}(\mathbf{x},t)$ is the
spatially filtered Eulerian fluid velocity in Eq. (\ref{g-star}).
Note that the continuity equation is now imposed as
$\nabla\cdot\hat{\mathbf{v}}= \nabla\cdot(g*\mathbf{v})=0$.
If filtering commutes with differentiation, then
$g*(\nabla\cdot\mathbf{v})$ also vanishes and the transported velocity
$\mathbf{v}$ is also divergenceless. The energy balance relation derived
from the NS$-\alpha$ equation (\ref{NS-alpha-eqns}) is
\begin{equation}\label{KE-alpha-diss}
\frac{d}{dt}\int
\frac{1}{2}\,\hat{\mathbf{v}}\cdot\mathbf{v}\ d^3x
=
\int
\hat{\mathbf{v}}\cdot\mathbf{f}\ d^3x
-
\nu \int
 {\rm tr}(\nabla\hat\mathbf{v}^T
 \cdot\nabla\mathbf{v})\ d^3x
\,,
\end{equation}
where we have used incompressibility and have dropped any boundary
terms that appear upon integrating by parts.


\paragraph{Vortex transport and stretching.} Let
$\mathbf{q}=\nabla\times\mathbf{v}$ be the vorticity of the
unfiltered Eulerian fluid velocity. The curl of the Kelvin-filtered
Navier-Stokes equation (\ref{NS-alpha-eqns}) gives the vortex
transport and stretching equation,
\begin{equation}\label{NS-alpha-vortex}
\frac{\partial\mathbf{q}}{\partial t}
+ \hat{\mathbf{v}}\cdot\nabla\mathbf{q}
- \mathbf{q}\cdot\nabla\hat{\mathbf{v}}
= \nu\nabla^2\mathbf{q}
+ \nabla\times\mathbf{f}\,.
\end{equation}
We note that the coefficient $\nabla\hat{\mathbf{v}}$ in the vortex
stretching term is the gradient of the {\bfi spatially filtered}
Eulerian velocity $\hat{\mathbf{v}}$. Thus, Kelvin-filtering tempers
the vortex stretching in the modified Navier-Stokes equation
(\ref{NS-alpha-eqns}), while preserving the original form of the
vortex dynamics. This tempered vorticity stretching is reminicent
of Leray's approach in regularizing the Navier-Stokes equations. Except
for the term $(\nabla\hat{\mathbf{v}})^{\rm T}\cdot\mathbf{v}$, the
Kelvin-filtered Navier-Stokes equation (\ref{NS-alpha-eqns}) is
otherwise quite similar to Leray's regularization of the Navier-Stokes
equations proposed in 1934 \cite{Leray[1934]}. Extension of the Leray
regularization to satisfy the Kelvin circulation theorem was cited as an
outstanding problem in Gallavotti's review \cite{Gallavotti[1992]}.
Indeed, for the case that $\mathcal{O}$ is the Helmholtz operator, full
advantage of this regularized vorticity stretching effect was taken in
\cite{FHT-JDDE[2000a]} to prove the global existence and uniqueness of
strong solutions for the three dimensional NS$-\alpha$ model. Paper
\cite{FHT-JDDE[2000a]} also proves the estimate (\ref{frac-dim-alpha})
for the fractal dimension of the global attractor for the NS$-\alpha$
model.


The {\bfi Navier-Stokes-alpha model} emerges from the
Kelvin-filtered Navier-Stokes equation (\ref{NS-alpha-eqns}) when
one chooses the operator $\mathcal{O}$ to be the Helmholtz
operator, thereby introducing a constant $\alpha$ that has
dimensions of length,
\begin{equation}\label{alpha-op}
\mathcal{O}
=
1-\alpha^2\nabla^2\,,
\quad\hbox{with}\quad
\alpha=const
\,.
\end{equation}
In this case, the filtered and unfiltered fluid velocities in
Eq. (\ref{g-inv}) are related by
\begin{equation}\label{g-inv-alpha}
\mathbf{v}=(1-\alpha^2\nabla^2)\hat{\mathbf{v}}
\,,
\end{equation}
and the reconstructed pressure $p$ is related to the hydrodynamic
pressure $P$ as

\begin{equation}\label{rec-pressure}
p=P -{1 \over 2} |\hat{\mathbf{v}}|^2 -
{{\alpha^2} \over 2}|\nabla\hat{\mathbf{v}}|^2
\,.
\end{equation}

The original derivation of the ideal {\bfi Euler-alpha model} (the
$\nu=0$ case of the NS$-\alpha$ model) was obtained by using the
Euler-Poincar\'e approach in \cite{HMR-98a}, \cite{HMR-98b}.
The physical interpretations of $\hat{\mathbf{v}}$ and $\mathbf{v}$ as
the Eulerian and Lagrangian mean fluid velocities were given in
\cite{DDH-fluct[1999]}.

The kinetic energy norm corresponding to (\ref{KE-norm}) for the
NS$-\alpha$ model is given by
\begin{equation}\label{KE-alpha}
E_\alpha = \int  \bigg[
\frac{1}{2}|\hat{\mathbf{v}}|^2
+ \frac{\alpha^2}{2}|\nabla\hat{\mathbf{v}}|^2 \bigg] d^3x
\,.
\end{equation}
This kinetic energy is the sum of a translational kinetic energy based on
the spatially filtered Eulerian velocity $\hat{\mathbf{v}}$, and a
gradient-velocity kinetic energy, multiplied by $\alpha^2$. By
showing the global boundedness, in time, of the kinetic energy
(\ref{KE-alpha}) one concludes that the coefficient
$\nabla\hat{\mathbf{v}}$ in the vortex stretching relation
(\ref{NS-alpha-vortex}) is {\it bounded in} $L^2$ for the NS$-\alpha$
model. Thus, the second term in the  kinetic energy (\ref{KE-alpha})
imposes an energy penalty for creating small scales. The spatial
integral of $|\nabla\hat{\mathbf{v}}|^2$ in the second term has the
same dimensions as filtered enstrophy (the spatial integral of
$|\nabla\times\hat{\mathbf{v}}|^2$, the squared filtered vorticity).
For a domain with boundary, or when considering transformations to
rotating frames, the spatial integral of
$\frac{1}{2}|\nabla\hat{\mathbf{v}}|^2$ in
the second term in the kinetic energy norm (\ref{KE-alpha}) should
be replaced by the integral of
${\rm trace}\,(\hat{\mathbf{e}}\cdot\hat{\mathbf{e}})$, where
$\hat{\mathbf{e}}$ is the strain rate tensor,
$\hat{\mathbf{e}}=(1/2)(\nabla\hat{\mathbf{v}}
+\nabla\hat{\mathbf{v}}^{\rm T})$
of the filtered velocity in Euclidean coordinates
\cite{Shkoller-preprint}. For the case we treat here, in Euclidean
coordinates and in the absence of boundaries, all three of these norms
are equivalent for an incompressible flow.


\section{Comparison of the NS$-\alpha$ model with LES equations}

The large eddy simulation (LES) equations for an incompressible flow are
conventionally  written as
\begin{equation}
\frac{\partial }{\partial t}\, \overline{u}_i +
\frac{\partial }{\partial x_j}\,\overline{u}_i\;\overline{u}_j   =
-  \frac{\partial} {\partial x_i}\, \overline{P}
+\nu \frac{\partial^2}{\partial x_j \partial x_j}\, \overline{u}_i
- \frac{\partial } {\partial x_j}\, \tau_{i\,j}
\label{eq:leseq}
\end{equation}

\begin{equation}
\hbox{with}\qquad
\frac{\partial\, \overline{u}_i}{\partial\, x_i} =  0.
\label{eq:contin}
\end{equation}
The overbar denotes spatial filtering which, for a quantity
$f(\mathbf{x})$, is defined by the integral

\begin{equation}
\overline{f}(\mathbf{x})
=
\int G(\mathbf{x}-\mathbf{x'}\,)\, f(\mathbf{x'}\,)\, d^3{x'}
\equiv
L_G{f}(\mathbf{x})
\,,
\label{eq:filter}
\end{equation}
where $G$ is a given kernel function, not necessarily the same as $g$
in Eq. (\ref{g-star}).  The quantities in Eq. (\ref{eq:leseq}), $u_i,\
P,$ and $\nu$ are the velocity components, pressure, and the kinematic
viscosity, respectively, and the effects of subgrid-scale quantities on
the resolved velocity are described by a subgrid-scale (SGS) stress
tensor

\begin{equation}
\tau_{i\,j}=\overline{{u}_i {u}}_j -
\overline{u}_i \; \overline{u}_j,
\label{eq:tau}
\end{equation}
which must be modeled in terms of the resolved quantities to close
Eq. (\ref{eq:leseq}). Note that the unfiltered velocity $u_i$ is
assumed to be a solution of the incompressible Navier-Stokes equation.

In contrast, the NS$-\alpha$ equation (\ref{NS-alpha-eqns}) with
definition (\ref{g-inv-alpha}) and (\ref{rec-pressure}) is written
in index notation as
\begin{equation}
\frac{\partial }{\partial t} v_i
+
\hat{v}_j \frac{\partial v_i }{\partial x_j}
+
v_k \frac{\partial \hat{v}_k }{\partial x_i}
=
-\,
\frac{\partial} {\partial x_i}
\left(
P - {1 \over 2} \hat{v}_k \hat{v}_k
-
{\alpha^2 \over 2}
\frac{\partial \hat{v}_m }{\partial x_l}
\frac{\partial \hat{v}_m }{\partial x_l}
\right)
+
\nu \frac{\partial^2}{\partial x_j \partial x_j} v_i,
\label{eq:alpha}
\end{equation}
where $\hat{v}_i$ is the filtered quantity related to the NS$-\alpha$
velocity $v_i$ through the equation
\begin{equation}
v_i= \hat{v}_i
-
\alpha^2 \frac{\partial^2}{\partial x_j \partial x_j} \hat{v}_i
=
\hat{v}_i - \alpha^2 \nabla^2 \hat{v}_i
\label{eq:hatv}
\end{equation}
and $\alpha$ is the filter length scale. Both fields $v_i$ and
$\hat{v}_i$ are incompressible since the `hat' filter commutes with
differentiation for a constant alpha.

The NS$-\alpha$ equation (\ref{eq:alpha}) can be rearranged into a form
similar to the LES equation (\ref{eq:leseq})

\begin{equation}
\frac{\partial }{\partial t} v_i +
\frac{\partial}{\partial x_j} v_i {v}_j  =
-\,\frac{\partial P} {\partial x_i} +
\nu \frac{\partial^2}{\partial x_j \partial x_j} v_i -
\frac{\partial}{\partial x_j}
\left(
(\hat{v}_j-v_j)v_i -\alpha^2
\frac{\partial \hat{v}_k }{\partial x_j}
\frac{\partial \hat{v}_k }{\partial x_i}
\right).
\label{eq:alpha1}
\end{equation}


\paragraph{Two interpretations of the NS$-\alpha$ velocity.} Both
the NS$-\alpha$ velocity $v_i$ and the {\sl filtered}
NS$-\alpha$ velocity $\hat{v}_i$ can be represented
numerically using a lower resolution than required to
represent the full NS velocity for the same problem. This is
the property that they share with the filtered velocity in
traditional LES. Nevertheless, it must be recognized from the
derivation in the previous section that there
is no LES filtering procedure that would produce the
NS$-\alpha$ equations from the full NS equations. Nevertheless
because of the similar behavior both NS$-\alpha$
velocities can be formally interpreted as the traditional LES
velocity.
If the NS$-\alpha$ velocity $v_i$ in equation (\ref{eq:alpha1}) is
interpreted as the filtered Navier-Stokes velocity $\overline{u}_i$ in
the LES equation (\ref{eq:leseq}) then the subgrid-stress tensor
associated with the alpha model is

\begin{equation}
\tau^\alpha_{i\,j}=
(\hat{v}_j-v_j)v_i -\alpha^2
\frac{\partial \hat{v}_k }{\partial x_j}
\frac{\partial \hat{v}_k }{\partial x_i}=
-\,\alpha^2 \left(
\frac{\partial \hat{v}_k }{\partial x_j}
\frac{\partial \hat{v}_k }{\partial x_i}-
\hat{v}_i \nabla^2 \hat{v}_j \right) -
\alpha^4 (\nabla^2 \hat{v}_i)(\nabla^2 \hat{v}_j)
\label{eq:tau_alpha1}
\end{equation}
where the last equality is obtained using Eq. (\ref{eq:hatv}). This is a
straightforward identification of terms.

Alternatively, one may propose to interpret the {\sl filtered}
NS$-\alpha$ velocity $\hat{v}_i$ as the filtered Navier-Stokes velocity
$\overline{u}_i$. To obtain the corresponding SGS stress in this case we
first filter the NS$-\alpha$ equation (\ref{eq:alpha1}) with the `hat'
filter. Since this filtering commutes with differentiation
one finds

\begin{equation}
\frac{\partial }{\partial t} \hat{v}_i
+
\frac{\partial}{\partial x_j} \hat{v}_i \hat{v}_j
=
-\frac{\partial \hat{P}} {\partial x_i}
+
\nu \frac{\partial^2}{\partial x_j \partial x_j} \hat{v}_i
-
\frac{\partial}{\partial x_j}
\left(
\Big(\widehat{\hat{v}_i \hat{v}_j} -\hat{v}_i \hat{v}_j\Big)
-
\alpha^2\bigg(
\widehat{\frac{\partial \hat{v}_k }{\partial x_j}
\frac{\partial \hat{v}_k }{\partial x_i}}
+
\widehat{\hat{v}_j \nabla^2 \hat{v}_i}\bigg)
\right)
\,,
\label{eq:alpha2}
\end{equation}
which involves doubly filtered quantities. The corresponding SGS stress
tensor is

\begin{eqnarray}
\tau^{f\alpha}_{i\,j}
\!\!& = &\!\!
(\widehat{\hat{v}_i \hat{v}_j} - \hat{v}_i \hat{v}_j)
-
\alpha^2\bigg(\widehat{\frac{\partial \hat{v}_k }{\partial x_j}
\frac{\partial \hat{v}_k }{\partial x_i}}
+
\widehat{\hat{v}_j \nabla^2 \hat{v}_i}\bigg)
\\
& = & (\widehat{\hat{v}_i \hat{v}_j}
-
\hat{\hat{v}}_i \hat{\hat{v}}_j)
-
\alpha^2 \widehat{\frac{\partial \hat{v}_k }{\partial x_j}
\frac{\partial \hat{v}_k }{\partial x_i}}
-
\alpha^2 \Big( \widehat{\hat{v}_j \nabla^2 \hat{v}_i}
-
\hat{\hat{v}}_j \nabla^2 \hat{\hat{v}}_i \Big)
\label{eq:tau_alpha2}
\\&&
+\,
\alpha^2 \hat{\hat{v}}_i \nabla^2 \hat{\hat{v}}_j
-
\alpha^4 (\nabla^2 \hat{\hat{v}}_i)(\nabla^2 \hat{\hat{v}}_j),
\nonumber
\end{eqnarray}
where the last equality is obtained using Eq. (\ref{eq:hatv}). We shall
next consider the transformation properties of the NS$-\alpha$ model. In
section \ref{sec-LES} we shall discuss how the SGS stress tensor
$\tau^{f\alpha}_{i\,j}$ for the filtered NS$-\alpha$ model is related to
LES similarity models for NS.


\section{Transformation properties}


\paragraph{Galilean invariance.}
Because of the first term in Eq. (\ref{eq:tau_alpha1}), or equivalently
the expression $\hat{v}_i \nabla^2 \hat{v}_j $ in the second part of
(\ref{eq:tau_alpha1}), the SGS tensor $\tau^\alpha_{i\,j}$ is neither
symmetric nor Galilean invariant, in contrast to the ordinary
SGS stress tensor (\ref{eq:tau}). However, the SGS force which is the
divergence of the SGS stress tensor is Galilean invariant for both
representations of $\tau^\alpha_{i\,j}$. Therefore, the dynamics of the
alpha model is Galilean invariant. Similarly, the SGS stress tensor
${\tau}^{f\alpha}_{i\,j}$ is not symmetric and lacks the Galilean
invariance because of the term
$\hat{\hat{v}}_i \nabla^2 \hat{\hat{v}}_j$ but the Galilean invariance
is restored in its divergence since $\hat{\hat{v}}_j$ is incompressible.


\paragraph{Rotating frame.}
It is of interest to consider a transformation of the SGS stress tensor
to a frame of reference rotating with a uniform angular velocity
$\Omega_i$. The full SGS stress tensor (\ref{eq:tau}) transforms in such
a way that the SGS force, i.e. $\partial \tau_{i\,j}/\partial x_j$, in an
inertial and in a  rotating frame are the same. Horiuti~\cite{horiuti00}
has recently analyzed several SGS models and showed that some of them do
not satify this condition.

The velocity in a rotating frame, denoted by the asterisk,
and the velocity in an inertial frame are related as follows

\begin{equation}
u_i = u^*_i + \epsilon_{imn}\Omega_m x^*_n
\,.\label{eq:uomega}
\end{equation}
Transforming the substantial derivative to a rotating frame and
using (\ref{eq:uomega}) gives

\begin{equation}
\frac{\partial }{\partial t} {u}_i +
\frac{\partial }{\partial x_j}{u}_i {u}_j =
\frac{\partial }{\partial t} {u}^*_i +
\frac{\partial }{\partial x^*_j}{u}^*_i {u}^*_j
+2\epsilon_{imn}\Omega_m u^*_n
+\epsilon_{imn}\epsilon_{nkl} \Omega_m \Omega_k x^*_l,
\label{eq:transf}
\end{equation}
where the last two terms are the {\bfi Coriolis force and the centrifugal
force},  respectively. Transforming the NS$-\alpha$ equation
(\ref{eq:alpha1}) to a  rotating frame gives

\begin{equation}
\frac{\partial }{\partial t} {v}^*_i +
\frac{\partial }{\partial x^*_j}{v}^*_i {v}^*_j =
-2\epsilon_{imn}\Omega_m v^*_n
-\epsilon_{imn}\epsilon_{nkl} \Omega_m \Omega_k x^*_l
-\frac{\partial P} {\partial x^*_i} +
\nu \frac{\partial^2}{\partial x^*_j \partial x^*_j} v^*_i
-\frac{\partial }{\partial x^*_j} (\tau^{\alpha*}_{i\,j}
+
\tau^{c*}_{i\,j}),
\label{eq:alpha1*}
\end{equation}
where $\tau^{\alpha*}_{i\,j}$ is the expression (\ref{eq:tau_alpha1}) for
`star' quantities and the additional correction term appears

\begin{equation}
\tau^{c*}_{i\,j}=\alpha^2\left(\epsilon_{imn} \Omega_m x^*_n \nabla^2
\hat{v}^*_j-
\epsilon_{kmj} \Omega_m \frac{\partial \hat{v}^*_k } {\partial x^*_i}-
\epsilon_{kli} \Omega_l \frac{\partial \hat{v}^*_k } {\partial x^*_j}\right).
\label{eq:correction}
\end{equation}
The divergence of this correction stress tensor does not vanish; in fact,

\begin{equation}
\frac{\partial }{\partial x^*_j}\tau^{c*}_{i\,j}=
\alpha^2 \left(
2\epsilon_{imn} \Omega_m \nabla^2 \hat{v}^*_n-
\epsilon_{kmj} \Omega_m
\frac{\partial^2 \hat{v}^*_k } {\partial x^*_i \partial x^*_j}
\right).
\label{eq:div_corr}
\end{equation}
However, the first term in (\ref{eq:div_corr}) combines with the
Coriolis force in Eq. (\ref{eq:alpha1*}) and the second term
combines with the centrifugal force to yield the following form of the
NS$-\alpha$ equation in the rotating frame:

\begin{equation}
\frac{\partial }{\partial t} {v}^*_i +
\frac{\partial }{\partial x^*_j}{v}^*_i {v}^*_j =
-2\epsilon_{imn}\Omega_m \hat{v}^*_n
-\frac{\partial \Pi} {\partial x^*_i} +
\nu \frac{\partial^2}{\partial x^*_j \partial x^*_j} v^*_i
+\frac{\partial }{\partial x^*_j} \tau^{\alpha*}_{i\,j},
\label{eq:alpha2*}
\end{equation}
where the total pressure $\Pi$ is

\begin{equation}
\Pi= P + {1 \over 2} (\Omega_m x^*_m)(\Omega_l x^*_l)
-{1 \over 2} (\Omega_m \Omega_m)(x^*_l x^*_l)
-\alpha^2 \epsilon_{mjk} \Omega_m
\frac{\partial \hat{v}^*_k }{\partial x^*_j}
\,.
\label{eq:Pi}
\end{equation}
\medskip


\paragraph{Remarks.}
This slight pressure shift is of no importance for incompressible fluid
dynamics. Had we replaced 
$\frac{1}{2}|\nabla\hat{\mathbf{v}}|^2$ in the second term of the
kinetic energy norm (\ref{KE-alpha}) by
${\rm trace}\,(\hat{\mathbf{e}}\cdot\hat{\mathbf{e}})$, as mentioned
earlier, we would have removed this pressure shift at the cost of making
the motion equation and stress tensor slightly more complicated. We
shall decline this option here; however, it shall be {\sl required} in
the compressible case.

Note that in the transformed motion equation (\ref{eq:alpha2*}) the
Coriolis force is computed using {\sl filtered} velocity
$\hat{\mathbf{v}}^*$. This is the physical velocity of the fluid parcels
in the rotating frame.

One may also obtain the {\sl nondissipative part} of the NS$-\alpha$
motion equation written in a rotating frame as in (\ref{eq:alpha2*}) by
first substituting
$\hat{v}_i = \hat{v}^*_i + \epsilon_{imn}\Omega_m x^*_n$ into
Hamilton's principle in which the Lagrangian is given by the kinetic
energy (\ref{KE-alpha}) and then taking variations in the
Euler-Poincar\'e framework, as discussed in \cite{HMR-98b}.


\section{Relation to generalized similarity models}\label{sec-LES}

We shall discuss the relation of the NS$-\alpha$ model to other
SGS models in terms of the inversion of the filtering operation
(\ref{eq:filter}). In generalized similarity models for the NS equations
the full NS velocity ${u}_i$ in the expression for the SGS stress tensor
(\ref{eq:tau}) is obtained by an inversion of the filtering operation
(\ref{eq:filter}), sometimes called {\bfi deconvolution}. 
The deconvolution is frequently encountered in
image processing and its properties are discussed in many 
textbooks on this subject, e.g.~\cite{andrews77,bates86}. In general, 
the inversion of the continuous filter is a ill-conditioned and often a
singular problem precluding a unique solution. This difficulty leads 
to a concept of an approximate or regularized deconvolution such as
a pseudo-inverse through a singular value decomposition 
approach~\cite{andrews77}. In the context of subgrid-scale modeling
the concept
of deconvolution was discussed  previously by several authors.
Germano~\cite{germano86} introduced an exponential filter for which  the
filtering operation can be inverted exactly, i.e. the function $f(x)$ in
(\ref{eq:filter}) can be represented as a closed form expression in
terms of the filtered function $\overline{f}(x)$. Using this
representation one obtains a formal expression for the subgrid-scale
stress tensor as a function of the filtered field.
Germano~\cite{germano86} shows that for this filter the dominant
term in the expression for the subgrid scale stress has the form  of a
nonlinear velocity gradient model

\begin{equation}
\tau_{ij} \sim
\frac{\partial \overline{u}_i}{\partial x_k}
\frac{\partial \overline{u}_j}{\partial x_k}.
\label{eq:nlmodel}
\end{equation}
This form, considered first by Leonard~\cite{leonard74}, has been more
recently investigated by Liu {\em et al.}~\cite{menev94}, Borue and
Orszag~\cite{borue98},  Leonard~\cite{leonard97}, and Winckelmans {\em
et al.}~\cite{winck98}.  In addition to the exact deconvolution
procedure of Germano~\cite{germano86} and Leonard~\cite{leonard97} other
approaches have been proposed and utilized  recently.
Geurts~\cite{geurts97} constructs an {\bfi approximate inverse operator}
by requiring that polynomials up to a certain order are recovered exactly
from their filtered counterparts.  Stolz and Adams~\cite{adams99} use a
formal  power series expansion of the inverse operator and obtain the
approximate inverse by truncating the expansion. Similar methods are used
by Horiuti~\cite{horiuti00} leading to a multi-level filtered model. The
deconvolution is also an integral part of the SGS estimation model of
Domaradzki {\it et al.} \cite{Domaradzki-Saiki97,domar99a,domar00a}. In
the spectral space version of the estimation model an exact
deconvolution is  possible~\cite{Domaradzki-Saiki97,domar00a}. In the
physical space representation the approximate deconvolution is
accomplished by solving a sequence of tridiagonal
equations~\cite{domar99a}.

Here, following Stolz and Adams~\cite{adams99} and
Horiuti~\cite{horiuti00} we shall consider a formal inverse of the
linear operator $L_G$ in (\ref{eq:filter}) as a power series expansion

\begin{equation}
L^{-1}_G =(I-(I-L_G))^{-1} = I + (I-L_G) + (I-L_G)^2 + ...
\label{eq:Linverse}
\end{equation}
where $I$ is the identity operator. To first order in the expansion
(\ref{eq:Linverse}) the velocity $u_i$ is approximated as

\begin{equation}
u_i=L^{-1}_G * \overline{u}_i \approx \overline{u}_i +
(\overline{u}_i - \overline{\overline{u}}_i)
\,.
\label{eq:approx}
\end{equation}
The second term in the approximation (\ref{eq:approx}) is obtained by
filtering of the difference $(u_i - \overline{u}_i)$ which can be
written using the Taylor expansion as

\begin{eqnarray}
u_i - \overline{u}_i
&=&
\int \big( u_i\,(\mathbf{x})-u_i\,(\mathbf{x}+\mathbf{x}'\,)\,
\big)
\,G(\mathbf{x}'\,)\,d^3x'
\\
 &=&
\int \left(-{{\partial u_i} \over {\partial x_j}}\, x'_j -
{1 \over 2} {{\partial^2 u_i} \over {\partial x_k \partial x_m}}
\, x'_k x'_m + \dots \right)
G(\mathbf{x}'\,)\,d^3x' \approx -\,
{{\Delta^2} \over {24}} \nabla^2 {u}_i\, (\mathbf{x})
\label{eq:approx1}
\end{eqnarray}
where the derivatives are evaluated at $\mathbf{x}$ and the above
relation is valid for both the symmetric top hat filter or 
the Gaussian filter with the width $\Delta$ in each Cartesian direction.
Using this  relation the approximate inversion (\ref{eq:approx}) becomes

\begin{equation}
u_i \approx \overline{u}_i
-{{\Delta^2} \over {24}} \nabla^2 \overline{u}_i
\,,
\label{eq:approxu_i}
\end{equation}
upon again invoking commutation of filtering and differentiation.
By comparing (\ref{eq:approxu_i}) for the approximate inversion and
(\ref{eq:hatv}) for the definition of $\hat{v}_i$ in the NS$-\alpha$
model, we recognize that $v_i$ and $\hat{v}_i$ are related by the
approximate inverse (\ref{eq:approx}) with $\alpha^2=\Delta^2/24$.
We shall pursue this relation farther by comparing the NS$-\alpha$ model
with LES generalized similarity models.

Employing the approximate inverse (\ref{eq:approxu_i}) in the
expression for the SGS stress tensor (\ref{eq:tau}) one finds

\begin{eqnarray}
\tau^{sim}_{i\,j} & = &\overline{{u}_i {u}}_j
- \overline{u}_i \; \overline{u}_j
\approx
(\overline{\bar{u}_i \bar{u}_j}
-
\bar{\bar{u}}_i\bar{\bar{u}}_j)
\nonumber \\
 && 
-\  \alpha^2 \left[(\overline{\bar{u}_j \nabla^2
\bar{u}_i}-
 \bar{\bar{u}}_j \nabla^2 \bar{\bar{u}}_i) +
(\overline{\bar{u}_i \nabla^2 \bar{u}_j}   -
\bar{\bar{u}}_i \nabla^2 \bar{\bar{u}}_j)
\right]
\label{eq:tausimilar}\\
 && 
+\  \alpha^4 (\overline{\nabla^2\bar{u}_i \nabla^2\bar{u}_j}-
 \nabla^2\bar{\bar{u}}_i \nabla^2\bar{\bar{u}}_j)
\,.
\nonumber
\end{eqnarray}
The generalized similarity tensor $\tau^{sim}_{i\,j}$ in expression
(\ref{eq:tausimilar}) is symmetric and Galilean invariant. On comparing
the last equation with the expression for $\tau^{f\alpha}_{i\,j}$ in Eq.
(\ref{eq:tau_alpha2}), we find as the most prominent difference the
product derivative term in the filtered alpha-model equation

\begin{equation}
\tau^{f\alpha}_{i\,j} =
\tau^{\alpha sim}_{i\,j}-\alpha^2 \left(
\widehat{\frac{\partial \hat{v}_k }{\partial x_j}
\frac{\partial \hat{v}_k}{\partial x_i}}
-\widehat{\hat{v}_i \nabla^2 \hat{v}_j} \right)
-\alpha^4 \widehat{(\nabla^2\hat{v}_i)(\nabla^2 \hat{v}_j}) =
\tau^{\alpha sim}_{i\,j}+\widehat{\tau^{\alpha}_{i\,j}},
\label{eq:newtau}
\end{equation}
where  $\tau^{\alpha sim}_{i\,j}$ is expression (\ref{eq:tausimilar})
written in terms of $v_i$ and the `hat' filter, instead of $u_i$ and the
`bar' filter. Also $\widehat\tau^{\alpha}_{i\,j}$ is the `hat'-filtered
version of $\tau^{\alpha}_{i\,j}$ in Eq. (\ref{eq:tau_alpha1}).
According to Eq. (\ref{eq:newtau})  the SGS stress tensor for the
filtered NS$-\alpha$ equation (\ref{eq:alpha2}) is the sum of the
similarity stress tensor for NS and the filtered SGS stress tensor
(\ref{eq:tau_alpha1}) for the original NS$-\alpha$ equation
(\ref{eq:alpha1}). The last equality is the
{\bfi Germano identity}~\cite{germano91} and follows from the the
definitions of $\tau^{\alpha sim}_{i\,j}$, $\tau^{\alpha}_{i\,j}$, and
$\tau^{f\alpha}_{i\,j}$.

We note that the important new term appearing in all SGS
stress tensor expressions for the alpha-model, namely the product of
derivatives, is different from the similar product of derivatives in the
nonlinear models (\ref{eq:nlmodel}).

The NS-$\alpha$ model is reminiscent, but not equivalent to the
generalized similarity models. This is encouraging because it may avoid
difficulties encountered by the pure similarity models. One of the most
serious problems encountered by the pure similarity models is
insufficient SGS dissipation caused by neglecting in the modeling scales
which cannot be resolved on a given LES grid. This fact has been long
recognized, e.g. by Zhou {\em et al.}~\cite{zhou89}, and more recently
by Domaradzki and Loh~\cite{domar99a}. Indeed, consider the most
favorable situation when the filtering operation can be inverted
exactly, i.e. the function $f(x)$ in (\ref{eq:filter}) can be
represented as a closed form expression in terms of the filtered
function $\overline{f}(x)$. In such a case one immediately obtains an
expression for the subgrid-scale stress tensor as  a function of the
filtered field from the definition (\ref{eq:tau}). However, the apparent
simplicity of such an exact approach to subgrid-scale modeling is
deceptive. This is because in the procedure $f(x)$ and $\overline{f}(x)$
have the same spectral support since the same LES mesh is used to
represent both functions. Therefore, no modeling of the subgrid scales
is actually performed in the deconvolution procedure. As a result
the similarity SGS stress tensor is unable to account for the
dissipative  effects of the nonlinear interactions between the resolved
scales and the unknown, subgrid scales which cannot be represented on a
given LES mesh. This implies that a good SGS model should either
explicitly model subgrid scales or possess extra terms in
addition to the similarity part. The former approach is used in the SGS
estimation modeling~\cite{Domaradzki-Saiki97,domar99a,domar00a} and the
additional terms are encountered in the mixed
models~\cite{bardina83,piom88,zhang93} as well as in the NS-$\alpha$
model (\ref{eq:newtau}).


\paragraph{Section \ref{sec-LES} summary.} We have identified the
`hat'-filtered NS$-\alpha$ equations with a class of generalized LES
similarity models in which the SGS stress tensor possesses additional
terms involving products of derivatives, as in
$\widehat\tau^{\alpha}_{i\,j}$. Applying the approximate-inverse methods
of deconvolution to this class of models restores the Kelvin circulation
theorem to the deconvolved solution. The corresponding dynamics of the
deconvolved solution is the NS$-\alpha$ model, whose analytical and
numerical properties are summarized in \cite{FHT-PhysD[2000b]}.


\section{Spectral space interpretation}\label{sec-spectral}

The nonlinear term in the NS$-\alpha$ equation (\ref{eq:alpha1})
transforms to spectral  (Fourier) space as

\begin{equation}
F.T.\left[\frac{\partial}{\partial x_j} v_l({\bf x})
\hat{v}_j({\bf x}) \right]
= ik_j\int  v_{l}({\bf p}) \hat{v}_j({\bf k-p}) d^{3}p
= ik_j\int  v_{l}({\bf k-p}) \hat{v}_j({\bf p}) d^{3}p
\,.
\label{eq:nlterm}
\end{equation}
Here the spectral representation is indicated by the dependence on
wavenumbers. The incompressibility condition requires projection of
(\ref{eq:nlterm}) onto a plane normal to the vector ${\bf k}$ which is
accomplished by applying the tensor

\begin{equation}
P_{lm}({\bf k}) = (\delta_{lm}-k_{l}k_{m}/k^{2}).
\label{eq:Pij}
\end{equation}

In spectral space one has

\begin{equation}
v_j({\bf k})= \hat{v}_j({\bf k}) + \alpha^2 k^2 \hat{v}_j({\bf k}),
\label{eq:FTfilter}
\end{equation}

so the projection of the nonlinear term can be written as

\begin{equation}
{1 \over 2}i \left(
k_j P_{lm}\int d^{3}p\, \frac{v_j({\bf p})}{1+\alpha^2 p^2}
v_{m}({\bf k-p})
+
k_m P_{lj}\int d^{3}p\, v_{j}({\bf p})
\frac{v_m({\bf k-p})}{1+\alpha^2 |{\bf k-p}|^2}\right).
\label{eq:nlterm1}
\end{equation}

This expression allows us to interpret the alpha-model in terms of triad
interactions. Qualitatively, the factor $1/(1+\alpha^2 k^2)$ attenuates
amplitudes of modes with wavenumbers $k > 1/\alpha$  and leaves modes
with $k < 1/\alpha$ largely unaffected.

Consider first a mode with $k < 1/\alpha$. Its triadic interactions with
all other modes ${\bf p}$ and ${\bf q}= {\bf k}-{\bf p}$ for which
$p,q < 1/\alpha$ are only slightly affected by the attenuation factors
and so are essentially the same as for the Navier-Stokes dynamics. If
one of the wavenumbers $p,q$ is greater than $1/\alpha$ its effect on
the mode
${\bf k}$ is weakened by one of the attenuation factors in
(\ref{eq:nlterm1}). In particular, the highly nonlocal interactions of
the eddy viscosity type, $k < 1/\alpha$ and $p,q \gg 1/\alpha$, are
removed from the dynamics.

Consider now a mode with $k > 1/\alpha$. In this case all local triads
with $p,q \sim O(k)$ are attenuated and the nonlocal triads with $p,q
\gg 1/\alpha$ are again removed from the dynamics. However, not all
nonlocal triads affecting the mode ${\bf k}$ are neglected. Indeed, for
$k > 1/\alpha$ it is possible to have triads with $p<1/\alpha<k$ and $q
\sim O(k)$ and then at least one of the terms in (\ref{eq:nlterm1}) is
not directly affected by the attenuation factors. The NS$-\alpha$ model
thus retains {\sl nonlocal sweeping effects} of large scales (either
$p$, or $q$ $<1/\alpha<k$) acting on small scales $k > 1/\alpha$.


\paragraph{Section \ref{sec-spectral} summary.} The NS$-\alpha$ model is
equivalent to the Navier-Stokes dynamics for all triads with $k,p,q <
1/\alpha$, it consistently neglects the nonlocal interactions of the
eddy viscosity type for modes $p,q > 1/\alpha$, and retains nonlocal
sweeping effects of large eddies on small eddies ($p$ or $q$
$<1/\alpha<k$ ).


\section{Acknowledgements}

It is our pleasure to acknowledge instructive and stimulating
discussions on many of these matters with P. Constantin, G. Eyink, C.
Foias, B. J. Geurts, N. Sandham, K. Sreenivasan, E. Titi, C. Vassilicos
and V. Yakhot. We are grateful for hospitality at the UC Santa Barbara
Institute for Theoretical Physics where this work was initiated during
their Hydrodynamic Turbulence program in spring 2000. ITP is supported
by NSF. One of the authors (JAD) was also partially supported by
the NSF Grant No. CTS-9704728.


\end{document}